\documentstyle[aps,prl,multicol]{revtex}
\input{epsf}
\begin{document}
\draft
\title{The Building Blocks of Spatiotemporal Intermittency}
\author{Martin van Hecke$^*$ }
\address{Center for Chaos and Turbulence Studies, The Niels Bohr Institute,
Blegdamsvej 17, 2100 Copenhagen \O, Denmark.}

\date{\today} \maketitle
\begin{abstract} 
We obtain a family of uniformly propagating
hole-solutions to the complex Ginzburg-Landau equation
which differ from the well-known Nozaki-Bekki holes.
They describe the spatial organization
and velocity of the dominant local structures in the
spatiotemporal intermittent regime. We
discuss the phenomenology of these intermittent
states in terms of the properties of the local structures.
\end{abstract}


\begin{multicols}{2} 
A proper understanding of spatiotemporal chaos, i.e.,
deterministic chaos occurring in extended systems
that are driven from equilibrium, is lacking.
Since the number of effective degrees
of freedom diverges with the system size,
most of the tools developed for low-dimensional systems
are inapplicable. Moreover,
these tools do not provide a proper framework
to describe the spatial organization of extended chaos.
In many cases, 
the dynamical states appear to be built up from local, almost 
particle-like objects with well-defined dynamics and interactions 
(Fig. $\!$1b-d).
A description of spatiotemporal chaos in terms of
these structures is therefore desirable \cite{loctoglob}.

In this Letter we will investigate the local structures that appear in
the {\em spatiotemporal intermittent} regime of the complex 
Ginzburg-Landau equation (CGLE):

\begin{equation}
A_t = A + (1+ i c_1)A_{xx} - (1-i c_3) |A|^2 A~. \label{cgle}
\end{equation}
This amplitude equation describes pattern formation near 
a Hopf bifurcation and
has been applied to describe patterns occurring in, e.g.,
fluid-convection, Faraday waves, optical systems, 
chemical oscillations, and turbulent flow past a wake \cite{CH}.
As a function of the coefficients $c_1$ and $c_3$,
which are determined by the underlying physical problem,
behavior ranging from completely regular to strongly chaotic 
has been found \cite{chao1,phasediagram} (Fig. $\!$1a).

In the spatiotemporal
intermittent regime \cite{phasediagram,STI1},
  a plain wave attractor coexists with a chaotic 
attractor; most initial conditions evolve to the latter 
(Figs. $\!1$b-d).
The typical states consist of patches of plane waves, separated
by various ``holes'', i.e., local structures characterized by a depression
of $|A|$. 
Similar intermittent states have been reported for the damped
Kuramoto-Sivashinsky equation, coupled CGLE's,
Rayleigh-B\'enard convection
and the printer instability \cite{STI2ks,maxiSM2,STI2rb,STI2pr}.
It has been suggested that spatiotemporal intermittency should occur 
generally in the transition route from laminar to chaotic states,
and the phenomenology suggests a relation to directed percolation 
\cite{STI2ks,STI3}.

The local structures in the intermittent regime
can be divided into two groups depending on 
the wavenumbers $q_l$ and $q_r$ of the asymptotic waves they connect.
The quasi-stationary structures in Fig. $\!$1c and 1d have
$q_l \neq q_r$

and are related to the 
intensively studied Nozaki-Bekki holes \cite{NB1}.
However, the dominant local structures have velocities and
asymptotic wavenumbers that are incompatible with the Nozaki-Bekki 
holes \cite{phasediagram}. For example,
the holes shown in Fig. $\!$1b all have $q_l\!\approx\!q_r\!\approx 0$
since the plain waves (gray areas) have wavenumbers close to zero;
similar holes are the dominant structures in the bi-chaotic regime.
In the following we shall characterize these holes 
and their dynamical properties, and discuss their relevance for 
the chaotic states of the CGLE.

In Fig. $\!$2, $|A|$, the  complex phase ($\arg(A)$) and
the local wavenumber $q\!:=\!\partial_x \arg(A)$
for the left lower part of Fig. $\!$1b are shown.
The wavenumbers of the laminar patches are quite close to zero,
while the cores of the local structures are
characterized by a sharp phase-gradient (peak

in $q$) and a dip of $|A|$.  The holes propagate with a 
speed of $0.95\pm 0.1$ and either their phase-gradient spreads out 
and the hole decays, or the phase-gradient steepens and the 
hole evolves to a phase-slip.
As a first step in describing these {\em local structures} which
have a slowly evolving velocity and spatial structure, 
we will study the {\em coherent} structures,
i.e., structures with {\em fixed} spatial structure and velocity.

For the 1D CGLE, coherent structures 
have been described in a simple framework  \cite{saar}. 
By substituting an {\em ansatz}
for a uniformly propagating solution of the form
$A(x,t)\!=\!e^{-i \omega t} \hat{A}(\xi)$ into the CGLE (\ref{cgle})
($\xi\!:=\!x-vt$), we obtain 
a set of coupled first order ordinary differential 
equations (ODE's) 
\begin{mathletters}\label{ode}
\begin{eqnarray}
\partial_{\xi} a &=& \kappa a ~,  \\
 \partial_{\xi} z
&=& -z^2 +\frac{1}{1+i c_1}\left[ - 1 - i \omega + (1-i c_3)a^2
-v z\right] ~,
\end{eqnarray} \end{mathletters}\noindent
where $a\!:=\! |\hat{A}|$ and where the complex quantity
$z$ is defined as $\partial_{\xi} \ln(\hat{A})\!=:\!\kappa + i q$.
Equation (\ref{ode}b) is equivalent to two real valued equations, so
(\ref{ode}) can be seen as a 3D real-valued dynamical system \cite{saar}.
Plain waves correspond to 
fixed points of (\ref{ode}), and the hole solutions we are interested
in correspond to orbits connecting these fixed points. 
A rather complete study of the heteroclinic orbits which describe
for instance the Nozaki-Bekki holes ($q_l \neq q_r$) has been made \cite{saar}.
We are  here interested in local structures that have
$q_l\!=q_r$, i.e., {\em homoclinic} orbits of the ODE's (\ref{ode}). 

In general the ODE's (\ref{ode}) have $\omega$ and $v$ as free parameters,
but since the wavenumber in the laminar patches is approximately zero,
we demand $q_l\!=\!q_r\!=\!0$, which fixes $\omega\! =\! -c_3$. 
The fixed point at $(a,z)\!=\!(1,0)$ corresponds
to the $q\!=\!0$ plain waves and
has a 1D outgoing manifold and a 2D spiraling ingoing manifold (Fig. $\!$3a).
To create a homoclinic orbit, we have to connect these manifolds.
This amounts to satisfying only a single 
condition, and since we have one free parameter $(v)$,
we can expect a discrete set of homoclons for $q_l\!=\!q_r\!=\!0$.
Performing a simple numerical integration of (\ref{ode}) and
adjusting the free parameter $v$, we obtain
a homoclinic orbit for $v\approx 0.916$; the corresponding 
coherent structure (Fig. $\!$3b) will be referred to as a 
{\em homoclon}.
For $q_l\!=\!q_r\!\neq0$, i.e., $\omega\!\neq\!-c_3$
one can obtain similar homoclinic orbits,
so in fact there exists a one-parameter family of homoclons.

In Fig. $\!$3b the homoclons are compared to the local structures
in the intermittent regime. The slight deviation 
between this particular local structure and the 
homoclon is mainly due to the fact
that the local structure has a slowly evolving shape.
The longer the lifetime of the holes, the better the fit is
to the homoclons.
For nearby values of $c_1$ and $c_3$ one finds similar correspondences
between the $q\!=\!0$ homoclons and the local structures.

\normalsize

It is instructive to compare the homoclons with the  
Nozaki-Bekki holes \cite{NB1}.
The Nozaki-Bekki holes contradict naive counting arguments and have
been shown to be structurally unstable, while
the homoclons satisfy the counting arguments, and are structurally stable.
One can verify that for $c_1\!=\!0.6,c_3\!=\!1.4$, a Nozaki-Bekki hole
with  $q_l\!=\!0$ has 
$q_r\!\approx\! 0.837$ and a velocity of $1.673$, 
which is completely different from the local structures and the
homoclons.
In some regimes, in particular for negative $c_1$, the Nozaki-Bekki holes
are dynamically relevant, and as they are sources for waves with
$q\neq0$, we obtain grain boundaries between $q\neq0$
and $q\approx0$ waves (Fig. $\!$1c,d).
In the limit where $q_l\! =\!q_r$, the width of the Nozaki-Bekki holes diverges,
so we can only conclude that there are two distinct types of hole 
solutions: heteroclinic Nozaki-Bekki
holes and homoclons.

Homoclinic orbits of (\ref{ode}) describe at least two other cases.
Since the homoclinic orbit occurs as a saddle-focus loop, there are
multi-circuit loops and a limit-cycle 
for nearby values of $v$ or $\omega$ \cite{shilnikov}. 
The limit cycle collapses via a Hopf 
bifurcation on another fixed point of the ODE's (\ref{ode}) 
for sufficiently small $v$.
The (quasi)periodic states of the CGLE 
that have been called ``compression waves'' \cite{comp} 
are described by these limit-cycles.
Furthermore, the ``riding chaos'' recently reported in \cite{maxiSM1} 
has been described in the context of homoclinic orbits 
for the {\em phase} equation of the CGLE, and
these orbits also occur in the ODE's (\ref{ode}).
For example, the uniformly propagating structures
shown in Fig. 11 of \cite{maxiSM1} occur in a background of wavenumber
$\approx 0.038$ for $c1\!=\!1.75, c3\!=\!0.8$.

The ODE's for the full CGLE (\ref{ode}) admit a homoclinic
orbit for $v\!\approx\!0.557$,
that describes the riding chaotic states perfectly.
For the same asymptotic wavenumbers and values 
of the coefficients, there
exists also a homoclinic orbit for $v\!\approx\!1.124$; 
this orbit is similar to the homoclons and
cannot be obtained from the phase equations.
More detailed studies of the homoclinic orbits are underway 
\cite{garcia,STIcoup,bohr}.

We will now return to the homoclons which appear 
in the spatiotemporal intermittent regime
and study their stability. Following an isolated,
slightly perturbed homoclon,
one immediately finds that they are weakly unstable \cite{spec}.
In a similar fashion
 to the local structures found in the intermittent state,
they either slowly decay or grow out to a phase-slip (Fig $\!$4a-d).
In the phase-space of the CGLE (\ref{cgle})
we can think of the homoclons as {\em unstable} equilibria
separating plain waves and phase-slips.

In Fig. $\!$4a we show the decay of a slightly perturbed homoclon
(perturbation of order $10^{-6}$). 
The total phase-difference $\Delta\theta$ across the decaying homoclon 
is conserved since there are no phase-slips;
for $c_1\!=\!0.6$ and $c_3\!=\!1.4$, $\Delta\theta\!\approx\! 3.24$.
During the decay,
the wavenumber-peak, amplitude-dip and apparent velocity 
decrease; for long times, the dynamics crosses over to a slow phase-diffusion
by which $\Delta \theta$ is smeared out.
In Fig. $\!$4b, the evolution of
a slightly perturbed homoclon towards several phase-slips is shown. 
The wavenumber-peak and amplitude-dip slowly grow, and
at $t\!\approx \!117$ the first phase-slip occurs, from which
a typical spatiotemporal intermittent state nucleates.
Before this phase-slip, the wavenumber acquires a negative peak in order to
conserve the total phase-difference across the structure (Fig. $\!$4c).
Both these peaks diverge at the phase-slip event, and just 
after the phase-slip, the winding-number $\int dx q/2\pi$
has decreased by 1.
Therefore, the negative phase-bump that corresponds to
the new left moving hole, is quite steep (Fig. $\!$4d), and
this hole will quickly grow out
to a new phase-slip, from which a strong right moving hole is generated etc.
When we quench $c_1$ and $c_3$ in the direction of 
the transition to plain waves, these zigzag motions
of the holes become very dominant (Fig. $\!$5a).

The group-velocity for the plain waves is close to zero.
Therefore, in the co-moving frame of a right moving hole, 
the wave to the right has its group-velocity 
pointing inward and the wave to the left has it pointing outward; 
hence the homoclons are neither sources nor sinks.  
Suppose we have a homoclon with a positive phase-difference
$\Delta \theta$, moving to the right into
a plain wave with wavenumber $q_r$. 
Because they are unstable, their dynamical fate
depends strongly on $q_r$.
When $q_r>0$, this leads to the ``winding up'' of the homoclon 
(increase of 
$\Delta \theta$), and a phase-slip is generated
in which the total winding-number is decreased. The larger $q_r$ is,
the faster 
this process goes, and as a result, 
the total number of phase-windings in the system is driven to         
approximately zero.  When $q_r<0$, or equivalently,
when a left moving homoclon invades a state with positive $q$,
this leads to the ``winding down'' of the homoclon, which then decays.
Its associated  $\Delta \theta$ is smeared out and
this drives the wavenumbers in the laminar patches to zero.
The winding up and down is shown in Fig. $\!$5b, where we follow the
evolution of an isolated homoclon in a background state with wavenumber
0.05.
This strong sensitivity to the asymptotic waves has as
a consequence that the fluctuations
in the laminar patches of the intermittent regime
have a strong effect on the homoclons. 
Small wave-packets, resulting from decaying homoclons, can trigger
the decay or phase-slip of the propagating holes; this might result in
correlations in the intermittent state over long times and distances.

The homoclons have only {\em one} weakly unstable eigenmode.
Consider the example illustrated in Fig. $\!$6a.  
The initial condition here has 
$|A|\!=\!1$ and a triangular wavenumber profile.
In general, there is no reason to expect such an initial condition
to evolve to a homoclon, but by adjusting only 
{\em one} parameter in the initial condition, this is precisely
what happens; for the example shown in
Fig. $\!$6a, the height of the triangle was set to 0.437754 
while its width was 20.
Increasing the height or width of the initial wavenumber-blob leads
to a steepening hole like in Fig. $\!$4b,
while a decrease leads to a decaying hole like in Fig. $\!$4a.
The total phase-difference

across the triangular initial condition is
$8.76$, so a big positive wavenumber-packet is emitted from the homoclon.
We have also tried this procedure for other initial conditions, and found that
in general the adjustment of only one free parameter in the initial
conditions leads to the generation of a long living homoclon;
the dimension of the unstable manifold is therefore one.
The contraction towards this unstable manifold is rather fast.
This is reflected in the dynamics 
in the intermittent regime; the local structures evolve to a reasonable
approximation along the 1D unstable manifold of the homoclons.

To substantiate this claim, we considered the relation between
the values of the extrema of $q$ and
the corresponding local minima of $|A|$.
As shown in Fig. $\!$6b, these quantities
are strongly correlated. This indicates that
a one-parameter family of profiles of $A$ is dominant.
Note the increase of the density around values of $a$ and $q$
which correspond to the coherent homoclon; being an equilibrium state,
the dynamics spend a relative long time there.

In Fig. $\!$6c,d we collapsed the profiles 
which correspond to certain isolated minima of $|A|$.
The profiles shown 
in Fig. $\!$6c correspond to the coherent homoclon, while
the profiles in Fig. $\!$6d correspond
to states evolving towards a phase-slip.

This suggests a phenomenological model in terms of moving
``homoclon'' particles possessing an internal degree of freedom 
that parameterizes the location on the unstable manifold.
More detailed studies of this are underway \cite{bohr}.

In conclusion, we have described a new class of coherent solutions
which occur in the spatiotemporal intermittent regime 
and which are intimately connected to phase-slips; they are therefore
one of the prime local structures of the 1D CGLE.

It is a pleasure to acknowledge discussions with Tomas Bohr,
Lorentz Kramer, Wim van Saarloos, 
Emilio Hern\'andez-Garc\'ia and Maxi San-Miguel.
This work was supported by the
Netherlands Organization for Scientific Research (NWO).

\end{multicols}

\begin{figure}[tbp]
\caption{ (a) 
``Phase-diagram'' of the CGLE. 
For small $c_1$ and $c_3$ all initial
conditions evolve to plain waves. In the intermittent
regime, a plain wave attractor and a chaotic attractor coexist.
Beyond the full curve
$c_1c_3\!=\!1$, all plain waves are linearly unstable 
and all states are spatiotemporal chaotic.
At a zero of $A$ the complex phase 
is undefined and phase-slips occur (see Fig. $\!$2); 
the chaotic state is then called defect-chaos.
When $A$ has no zeroes we speak of phase-chaos.
In the bi-chaotic regime, a 
defect and phase-chaotic attractor coexist [4,11].
(b)-(d) Space-time plots (over a range of $200\times150$)
of $|A|$ (black corresponds to $|A|\!=\!0$)
showing chaotic states in the spatiotemporal intermittent regime, 
for coefficients
$(c_1,c_3)\!=\!$ $(0.6,1.4)$ (b), $(0,1.8)$ (c) and $(0,1.4)$ (d).}
\end{figure}

\begin{figure}[tbp]
\caption{Space-time (60x50) plots of the left-lower part of
Fig. $\!$1b, showing $|A|$, $\arg(A)$ and
$q\!:=\!\partial_x \arg(A)$ in detail.}
\end{figure}

\begin{figure}[tbp]
\caption{
(a): The homoclinic orbit of the ODE's (2) in $a,q,\kappa$ space.
(b) The amplitude and wavenumber profile of the
corresponding coherent structures (curves). 
The circles correspond to a local
structure obtained from simulations of the CGLE
in the spatiotemporal intermittent regime.
The bump in $a$ to the right of the core corresponds to
the spiraling motion on the incoming manifold of the fixed point
$(a,q,\kappa)\!=\!(1,0,0)$.}
\end{figure}

\begin{figure}[tbp]
\caption{(a,b) Evolution of the wavenumber profiles of slightly 
perturbed homoclons.  Consecutive time-slices 
have a time difference of 5.
(c,d) Wavenumber profiles (full curves) and 
winding-number $\int dx q/2\pi$ (dashed curves)
just before (c) and after (d) the first phase-slip. }
\end{figure}

\begin{figure}[tbp]
\caption{(a) The dominance of zigzagging holes near the transition to plain waves 
$c_1\!=\!0.6,c_3\!=\!1.2$ for space $\times$ time $= 512 \times 1000$.
(b) The evolution of a homoclon in a background state with wavenumber 0.05,
for $c_1\!=\!0.6,c_3\!=\!1.4$ and space $\times$ time $= 512 \times 250$.}
\end{figure}

\begin{figure}[tbp]
\caption{(a) The wavenumber profiles of a ``triangular'' initial condition evolving
to a homoclon. (b) Collapse of the extrema of $a$ and $q$ in the intermittent
regime. (c) Amplitude profiles of $|A|$ with minima of $|A|$ around the
homoclon value. (d) Similar profiles for steeper minima. }
\end{figure}


\begin{references}
\bibitem{salt}
The diagram should be viewed as a rough indication of typical
behavior since it is not always clear how to obtain or define the 
``transition'' curves.
\bibitem{STIcoup}
The spatiotemporal intermittent states that have been reported in coupled
CGLE's \cite{maxiSM2} also have propagating local structures.
It appears that these structures are related to
homoclinic orbits, but now for a set of 6 coupled
ODE's wwhich describe the coherent structures in coupled CGLE's.
\bibitem{spec} 
To study the stability properties of the homoclons, one
should calculate their spectrum; these calculations 
will be performed in the future \cite{bohr}.
\end{references}
\end{document}